# Mechanical Detuning of Exciton-Phonon Resonance in $WS_2$

Álvaro Rodríguez[1]†*, Carmen Munuera[1], Andres Castellanos-Gomez[1]

[1]Instituto de Ciencia de Materiales de Madrid (ICMM-CSIC), C. Sor Juana Inés de la Cruz, 3, Madrid, 28049, Spain.

† Departamento de Física de la Materia Condensada, Universidad Autonóma de Madrid and Condensed Matter Physics Center (IFIMAC), Universidad Autonóma De Madrid, Madrid 28049, Spain.

*Corresponding author(s): alvaro.rodriguez@csic.es

**TOC Graphic**

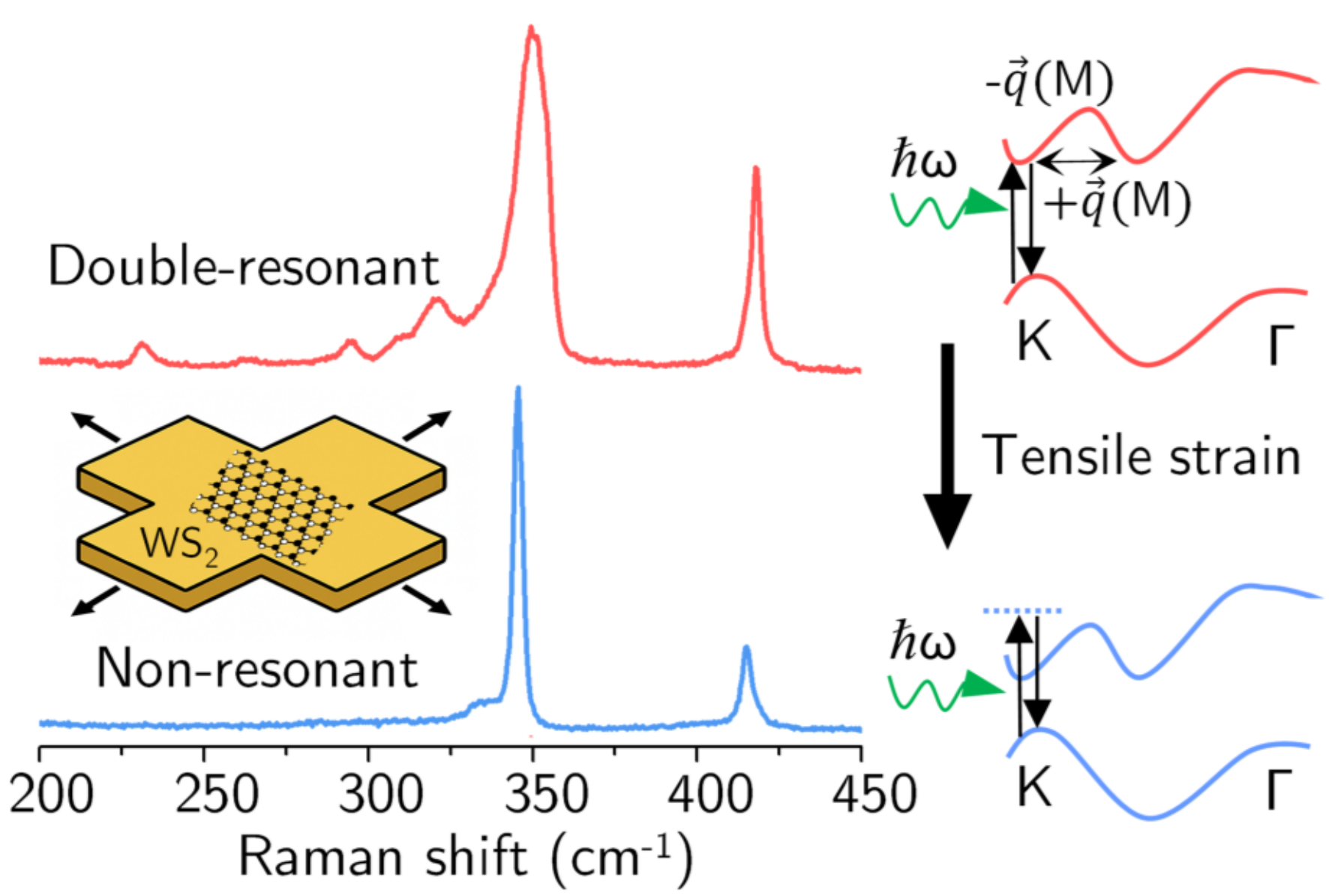

## ABSTRACT

Controlling resonant Raman scattering in two-dimensional semiconductors typically requires tuning the excitation energy to match excitonic transitions. Here we show that mechanical deformation can achieve the same effect without changing the laser energy, enabling a controlled transition between resonant and non-resonant Raman scattering at fixed excitation. By applying biaxial strain up to 1.3% to $WS_2$, the B exciton is red-shifted by 180 meV. This large excitonic shift leads to a pronounced collapse of the double-resonant 2LA(M) mode under 532 nm excitation, quantitatively described by a resonance model formulated in terms of the B exciton energy. Meanwhile, first-order phonons remain narrow and reversible, confirming elastic deformation and efficient strain transfer. These results establish mechanical strain as an effective knob to control exciton-phonon mediated light-matter interactions. They enable deterministic and reversible tuning of resonance-enhanced Raman scattering and excitonic optical responses in layered semiconductors.

## Introduction

Transition metal dichalcogenides (TMDs) combine tightly bound excitons with strong electron-phonon interactions, which makes them promising materials for optoelectronic and photonic technologies where optical responses can be dynamically modulated[1–3] Mechanical strain provides a powerful means to tune their electronic structure by shifting exciton energies, modifying bandgaps and altering valley properties.[4–9] Most strain-engineering studies have focused on uniaxial deformation, while the application of high biaxial strain has remained challenging because generating isotropic in-plane expansion over large areas is experimentally demanding.[10–13] Approaches based on bubbles, wrinkles, suspended membranes, patterned substrates or thermal expansion mismatch can produce substantial strain, but often over small or spatially non-uniform regions and with limited control over the resulting deformation field[14–20] Bending flexible substrates in a cruciform geometry offers a straightforward route to biaxial deformation over large areas. However, when flakes are transferred directly onto polymers, the relatively weak interfacial interaction can lead to partial strain transfer and sliding, restricting the strain effectively experienced by the material. These limitations have hindered quantitative studies of exciton-phonon coupling under biaxial strain. Earlier work has largely concentrated on photoluminescence shifts or bandgap renormalization[21,22] whereas the influence of biaxial strain on resonant Raman scattering has been limited to application of small strain values or inefficient strain transfer.[12,23,24] In particular, although previous studies have reported strain-induced modifications of resonant Raman features, no experimental work has demonstrated that excitonic detuning driven solely by strain can modulate double-resonant Raman processes, a mechanism that is typically controlled by tuning the excitation energy.

Recently, we showed that direct exfoliation of $MoS_2$ onto ultrathin gold films markedly improves strain-transfer efficiency and enables access to strain values beyond those achievable with conventional polymer-supported samples.[25] Building on this strategy, here we combine gold-assisted exfoliation with a cruciform bending platform to apply high and spatially uniform biaxial strain to $WS_2$ over regions of several hundred micrometers.[10] The presence of gold during exfoliation promotes stronger adhesion and suppresses sliding, enabling more efficient transfer of the applied deformation to the 2D crystal while preserving optical quality on Au/polycarbonate (PC) substrates.[25]

In this work, we demonstrate that biaxial strain continuously red-shifts the A and B excitons and mechanically detunes the system from resonant to non-resonant Raman scattering. The resulting suppression of the 2LA(M) band is accurately described by a resonance model expressed in terms of the exciton energy and the finite exciton-assisted scattering window. To date, no experimental study has demonstrated that biaxial strain alone can drive a controlled transition between resonant and non-resonant Raman regimes in a layered semiconductor. This approach establishes biaxial strain as a practical and reversible control knob to access resonant and non-resonant Raman regimes at a fixed excitation energy and provides a route to mechanically programmable Raman responses in van der Waals materials.

## Results and Discussion

### Experimental configuration and strain calibration

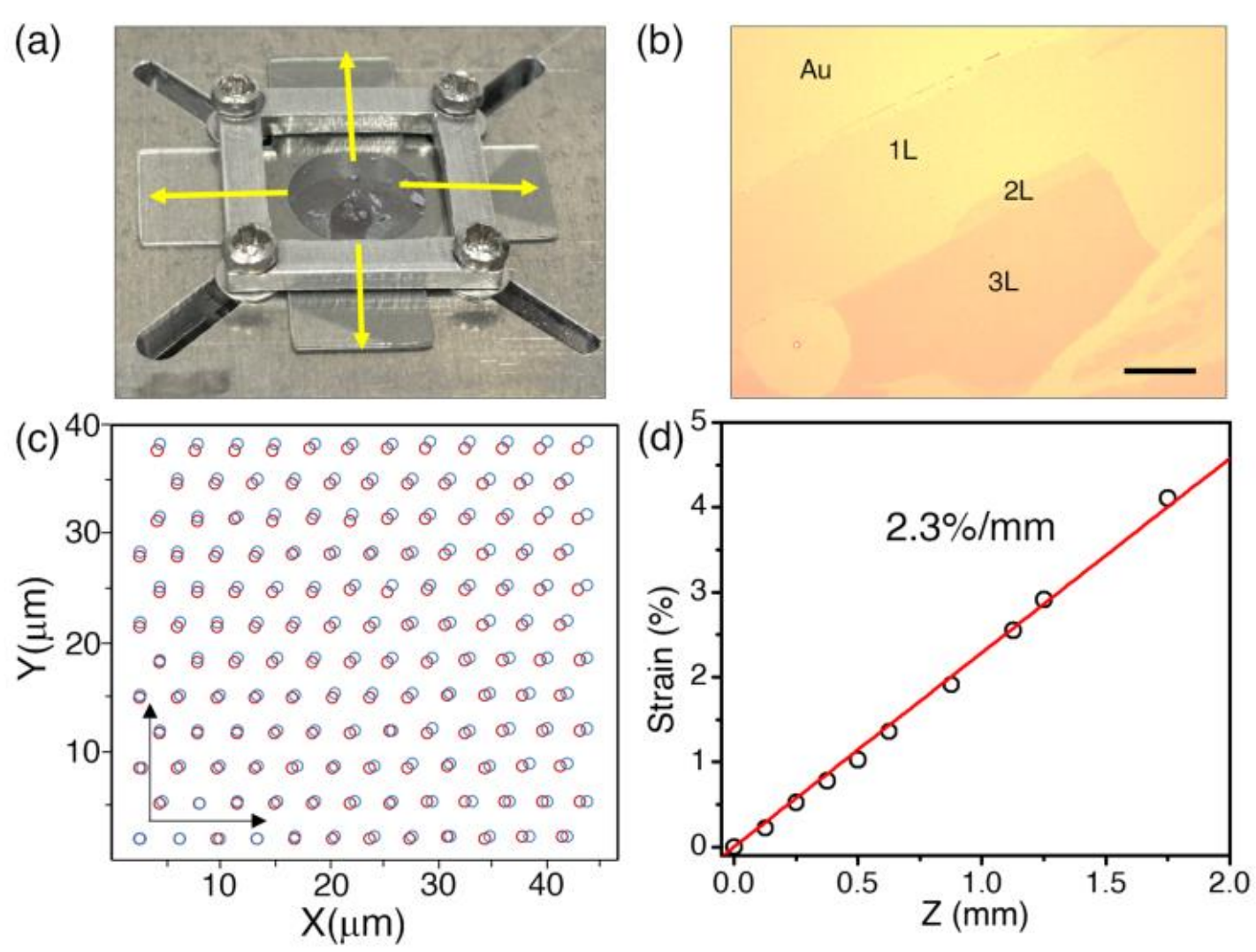

**Figure 1. Experimental configuration and strain calibration.** (a) Photograph of the cross-shaped flexible polycarbonate (PC) substrate used to apply biaxial strain. The central region hosts the large $WS_2$ layers and mechanical strain is applied along yellow crosses. (b) Optical micrograph of the $WS_2$ monolayer, bilayer and trilayer regions on Au/PC. Scale bar is 10 µm. (c) Pillar-array calibration of the biaxial strain. Optical images collected before and after bending at 2% strain reveal isotropic expansion of pillar spacing. (d) Strain extracted from pillar displacements as a function of Z-stage position (Z). The linear fit ($\varepsilon = kZ$) provides the calibration used throughout this work.

To impose biaxial strain, we use a cruciform bending geometry in which $WS_2$ is placed on the tensile side of a cross-shaped PC substrate (Figure 1(a)).[10,26,27] Out-of-plane displacement of the central stage produces isotropic in-plane expansion in the central region, allowing uniform biaxial deformation while minimizing shear. Large $WS_2$ flakes were obtained by gold-assisted exfoliation,[28–30] which yields clean interfaces and strong adhesion to the Au/PC substrate, ensuring efficient strain transfer to the $WS_2$ (Figure 1(b)).

Layer thickness was independently confirmed by AFM measurements and Raman spectroscopy (see Supporting Information, Figure S2). All samples in this work were prepared on Au, but the effect of the metal substrate is thickness dependent. In monolayer and bilayer $WS_2$, the proximity to the metal interface more strongly influences the electronic structure and optical spectra due to enhanced electronic hybridization and modified dielectric screening at the interface, leading to increased background and more intricate Raman features.[30,31] As thickness increases, the electronic and vibrational response becomes progressively less sensitive to the substrate, and the trilayer more closely reflects the intrinsic properties of $WS_2$ while maintaining efficient strain transfer. In addition, the trilayer tolerates higher biaxial strain while preserving well-defined spectral features, as the A and B excitons remain well resolved in differential reflectance under strain. For these reasons, trilayer $WS_2$ provides a particularly robust platform to quantitatively track the strain-driven excitonic detuning and resonance suppression. The underlying mechanism, however, is not restricted to a specific layer number.

Strain calibration was performed using a pillar-array reference patterned on an identical substrate. The biaxial strain, $\varepsilon = \Delta L/L_0$, was calculated from the relative change in the pillar spacing before and after bending (Figure 1(c)), and found to increase linearly with the vertical displacement of the Z-stage, $\varepsilon = kZ$ (Figure 1(d)). This calibration was used to convert all actuator displacements into absolute biaxial strain values. The isotropy of the strain field is confirmed by the linear softening of both the in-plane E mode and the out-of-plane $A_1$ mode. We adopt the E and $A_1$ notation corresponding to the $C_{3v}$ symmetry of $WS_2$ on Au[31]. All Raman measurements were performed at a fixed excitation energy (532 nm), such that changes in the Raman response directly reflect strain-induced excitonic detuning rather than excitation-energy tuning. Raman peak positions and intensities were obtained by multi-peak fitting of the measured spectra. Raman peaks were fitted using Voigt profiles. While the intrinsic phonon lineshape is expected to be predominantly Lorentzian, the experimental spectra were acquired under resonant conditions and are subject to finite instrumental resolution as well as possible inhomogeneous broadening over the optically probed region. The Voigt function therefore provides a more realistic description as a convolution of Lorentzian (intrinsic lifetime broadening) and Gaussian (instrumental and inhomogeneous) contributions.

To directly verify the strain uniformity within the $WS_2$ flake, Raman mapping was performed under applied biaxial strain. The resulting spatial strain distribution confirms a homogeneous deformation across the optically probed region (see Supporting Information, Figure S5).

**Resonant Raman response of trilayer $WS_2$**

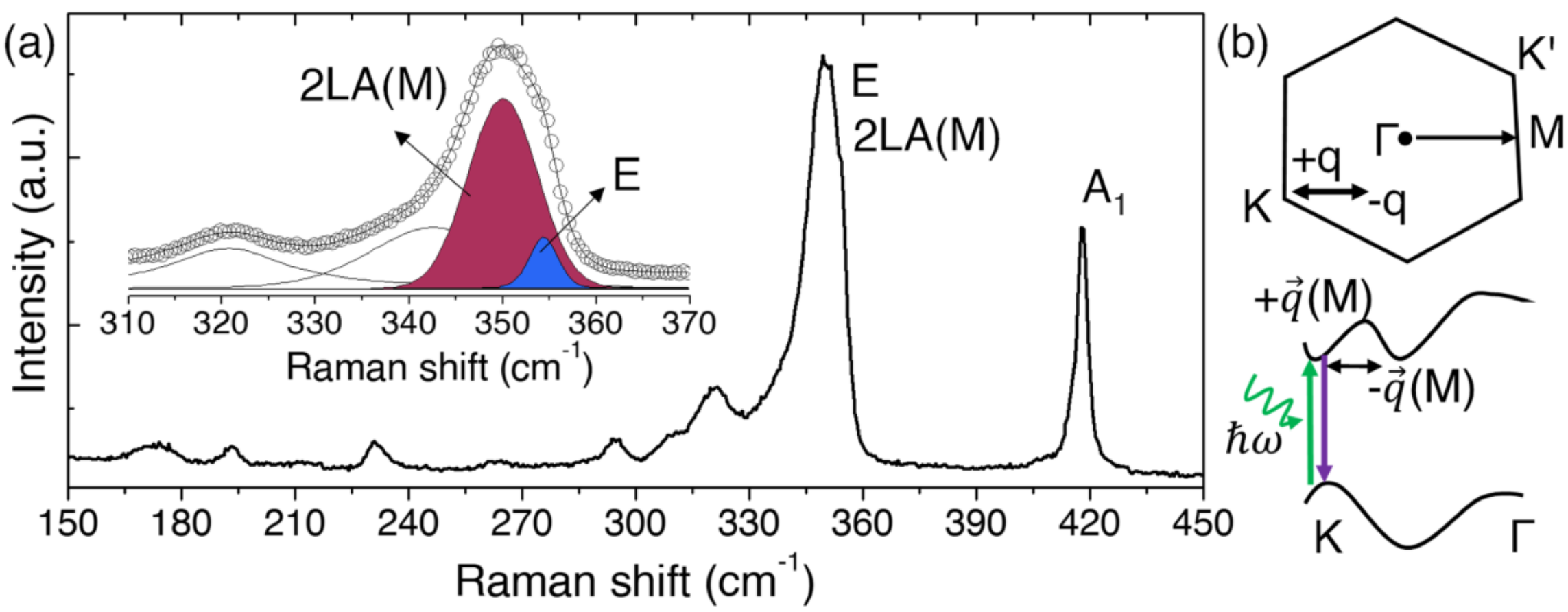

**Figure 2. Resonant Raman features in trilayer $WS_2$ under 532 nm excitation.** (a) Raman spectrum at zero strain ($\varepsilon = 0$ %) showing the resonant enhancement of the 2LA(M) band. The spectral decomposition highlights contributions from the E, $A_1$, and 2LA(M) modes obtained by multi-peak fitting (Voigt profiles). Peaks labels and color- match those used in the strain-dependent measurements. Overlap between the E phonon and the double-resonant 2LA(M) mode produces an asymmetric band whose intensity and shape are highly sensitive to exciton-mediated resonance. (b) Schematic representation of the double-resonant 2LA(M) Raman process. An optical excitation near K is followed by two intervalley scattering events mediated by LA phonons with wavevectors ±q close to M, before radiative recombination. This mechanism underlies the strong resonance sensitivity of the 2LA(M) feature.

Figure 2 shows the Raman spectrum of trilayer $WS_2$ under 532 nm excitation, which lies close to the B exciton energy and enhances both first-order (E and $A_1$) phonons and double-resonant processes.[32–35] The E phonon appears at ~355 $cm^{-1}$ and the $A_1$ phonon at ~417 $cm^{-1}$, accompanied by a strong 2LA(M) band.[36,37] The 2LA(M) mode arises from an intervalley double-resonant process involving intermediate electronic states near the excitonic transition. Its intensity is therefore highly sensitive to the proximity between the laser energy and the excitonic absorption. Its Raman shift corresponds to the energy of two longitudinal acoustic phonons with wavevector close to M. Although theoretical analyses show that double-resonant Raman processes in TMDs can involve scattering pathways in different regions of the Brillouin zone, including states near K, these contributions are not resolved as separate Raman bands in our spectra. The notation 2LA(M) is therefore used here in the conventional experimental sense, referring to the dominant phonon wavevector associated with the observed feature[36,38].

A schematic representation of this double-resonant scattering pathway is shown in the Figure 2(b), where the two LA phonons with opposite wavevectors connect electronic states across the Brillouin zone, making the 2LA(M) intensity highly sensitive to excitonic resonance. The partial overlap between E and 2LA(M) produces an asymmetric profile that is highly sensitive to exciton-mediated resonance and serves as a reference for monitoring the evolution of exciton-phonon coupling under biaxial strain.

**Transition from resonant to non-resonant Raman scattering under strain**

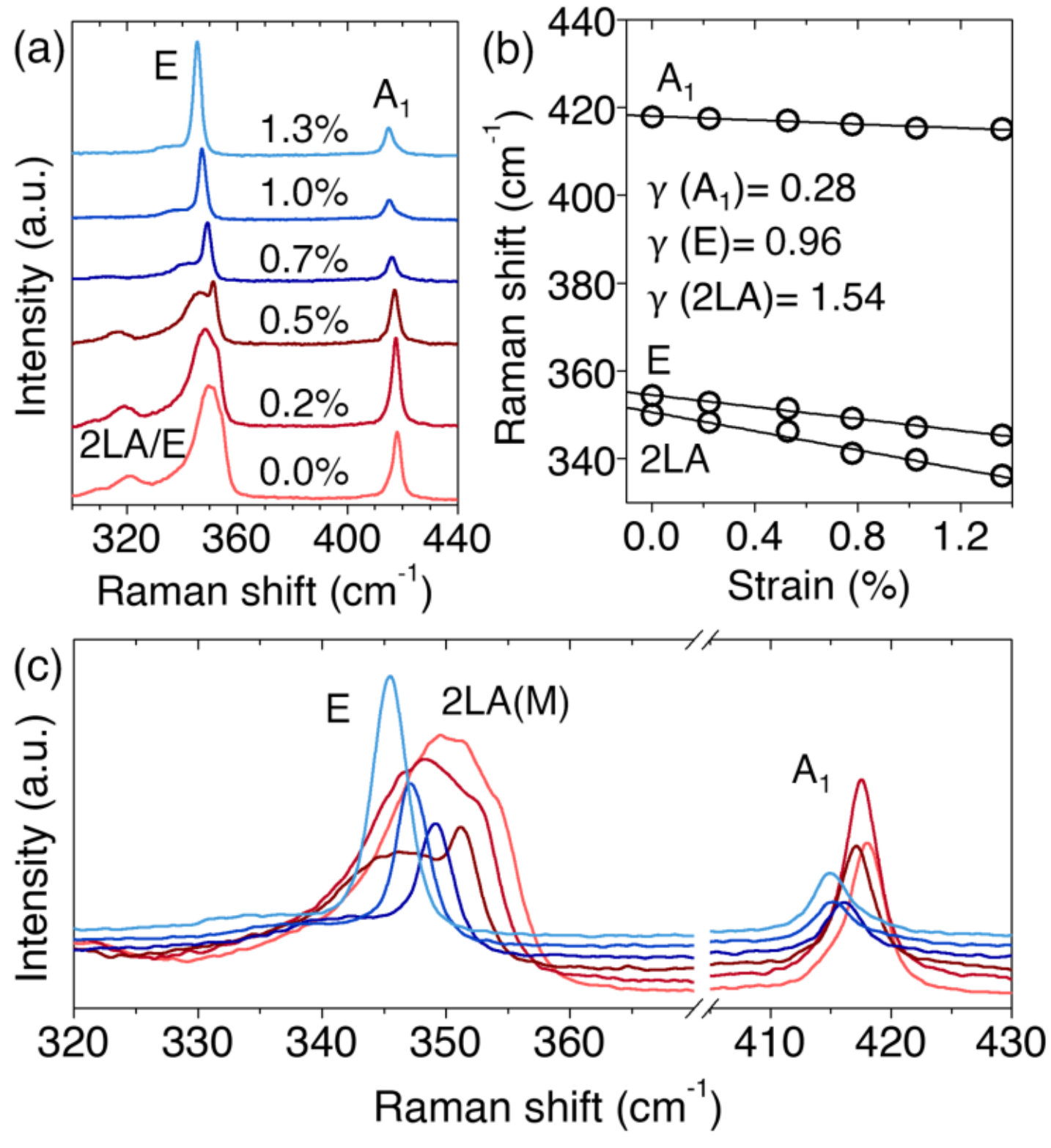

**Figure 3. Phonon softening and resonant Raman evolution in biaxially strained $WS_2$.** (a) Raman spectra of trilayer $WS_2$ acquired at increasing biaxial strain. (b) Strain dependence of the E, $A_1$, and 2LA(M) phonon frequencies. Linear fits yield slopes of -6.8 ± 0.10 $cm^{-1}$/% and -2.3 ± 0.08 $cm^{-1}$/%, and -10.8 ± 0.6 $cm^{-1}$/% respectively. Using $\gamma = -\frac{1}{2\omega_0}\left(\frac{d\omega}{d\varepsilon}\right)$, we obtain the Grüneisen parameter for each Raman mode. $\gamma(E)$=0.96 ± 0.05, $\gamma(A_1)$=0.28 ± 0.03 and $\gamma$(2LA)=1.54 ± 0.15. (c) Zoomed view of the spectra highlighting red-shifts and the reduction of resonant enhancement.

Figure 3(a) displays Raman spectra recorded from 0 to 1.3% strain. Both first-order phonons red-shift systematically, with the E mode showing the stronger dependence, reflecting its in-plane character and the efficiency of isotropic strain transfer.[12] A closer inspection (Figure 3(c)) highlights the simultaneous softening of the phonons and the reduction of the 2LA(M) enhancement as strain increases. The absence of mode splitting further confirms that the deformation remains effectively isotropic, in contrast to uniaxial strain where the E mode separates into two components.[9] Linear fits yield gauge factors of −6.8 ± 0.1 $cm^{-1}$/% for the E mode and −2.3 ± 0.08 $cm^{-1}$/% for the $A_1$ mode, consistent with previous reports for biaxially strained $WS_2$ (Figure 3(b)). Expressing these strain sensitivities in terms of the Grüneisen parameter[39,40], $\gamma = -\frac{1}{2\omega_0}\frac{d\omega}{d\varepsilon}$ under biaxial deformation, where $\omega_0$ is the phonon frequency at zero strain, yields $\gamma(E) = 0.96 \pm 0.05$ and $\gamma(A_1) = 0.28 \pm 0.03$, values consistent with efficient and isotropic strain transfer and with previously reported values for biaxially strained $WS_2$.[12,23]

Raman measurements for monolayer and bilayer $WS_2$ are provided in the Supporting Information (Figure S3). Comparable phonon strain sensitivities are observed in monolayer and bilayer $WS_2$, although the apparent shifts are generally smaller than in the trilayer. This difference should be interpreted in the context of substrate and resonance effects. In monolayer and bilayer $WS_2$ supported on Au, the layer in direct contact with the metal can experience additional interfacial interactions, including residual strain electronic hybridization, and modified dielectric screening, which increase spectral complexity and partially obscure intrinsic strain responses. As thickness increases, the influence of the metal is progressively screened, and the trilayer more closely reflects the intrinsic lattice response to externally applied strain. For the 2LA(M) feature, this distintion is further amplified by its double-resonant character. While a reduction of the 2LA(M) intensity is also observed in mono- and bilayer regions, substrate-related effects render the resonance condition less well defined, complicating a quantitative comparison. The trilayer therefore provides the clearest regime to track the strain-driven evolution of excitonic detuning and resonance suppression under the present experimental conditions.

Importantly, the larger strain window accessible in our samples, reaching up to 1.3% compared to the 0.5–0.7% typically reported in earlier studies, enables direct observation of the strain-induced suppression of the 2LA(M) mode, which was not resolved at lower strain levels.

**Exciton shifts under biaxial strain**

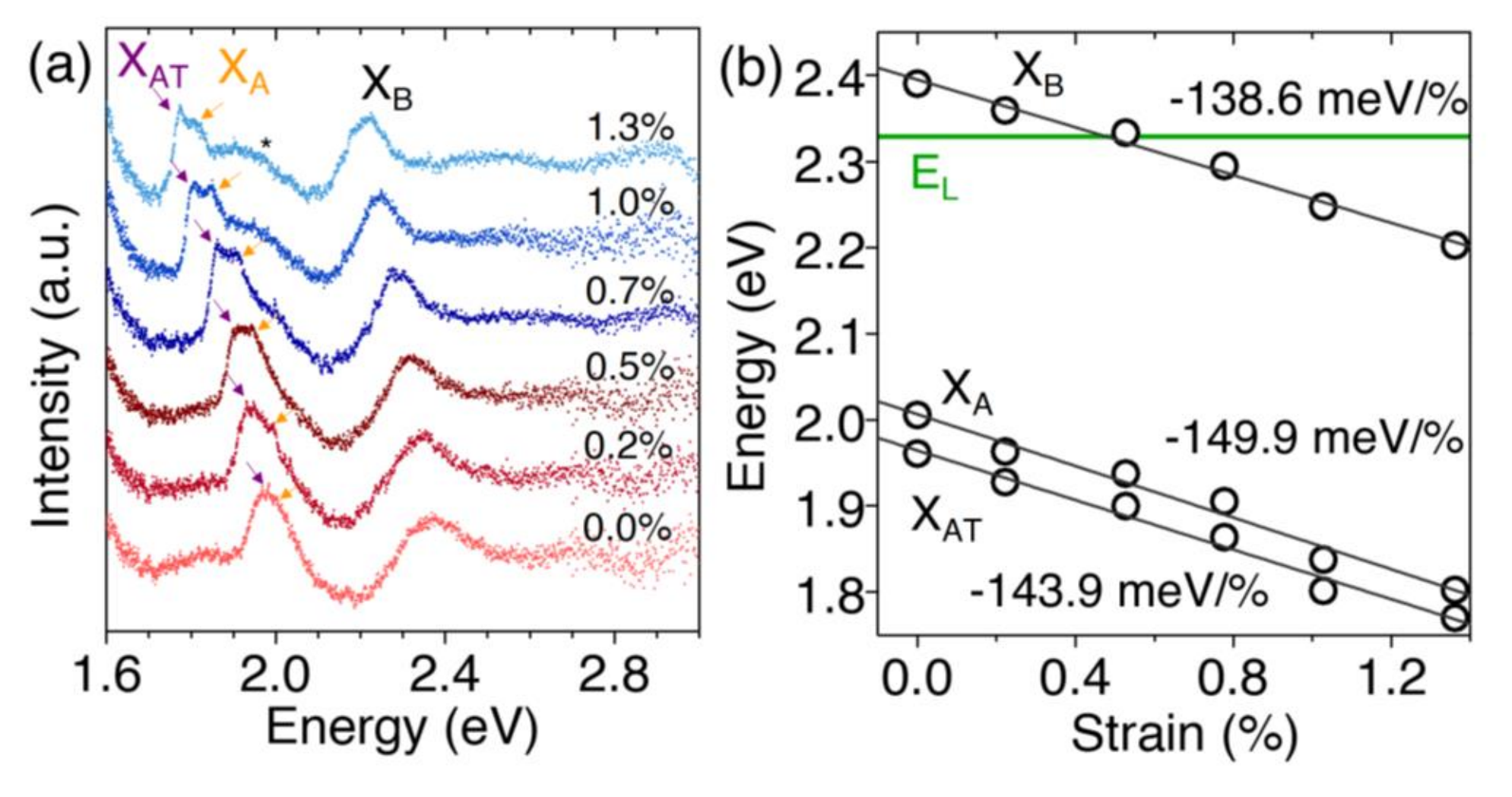

**Figure 4. Differential-reflectance tracking of exciton energies under biaxial strain.** (a) Normalized ΔR/R spectra showing the systematic red-shift of the A and B excitons with increasing tensile biaxial strain. Purple and orange arrows indicate the A trion ($X_{AT}$) and exciton ($X_A$) positions, respectively. The background signal was subtracted from each spectrum for clarity. (b) Strain dependence of the $X_A$, $X_B$, and $X_{AT}$ energies. Linear fits yield a slope of -138.6 ± 10 meV/% for the B exciton, corresponding to a total shift of ≈-180 meV at $\varepsilon$ = 1.3%. The green line marks the laser excitation energy ($E_L$).

Differential-reflectance spectra in Figure 4(a) show clear red-shifts of both A and B excitons with increasing biaxial strain. A representative example of the background subtraction procedure is provided in the Supporting Information (Figure S4). At high strain, the A exciton ($X_A$) develops a lower-energy shoulder attributed to the trion contribution ($X_{AT}$), [16,24] whereas the B exciton ($X_B$) remains unsplit across the full strain window. Additionally, a pronounced high-energy shoulder develops on the A-exciton feature at higher strain (marked by an asterisk). This feature may originate from strain-induced modifications of the excitonic fine structure, including changes in the trion–exciton balance, excited excitonic states, or strain-dependent dielectric screening.[41,42] A possible contribution from interlayer excitonic transitions, similar to effects reported in bilayer $MoS_2$, cannot be excluded.[43] The smooth and continuous evolution with strain indicates that it reflects intrinsic modifications of the excitonic landscape rather than a measurement artifact.

Linear fits in Figure 4(b) yield shift rates of −150 meV/% for the A exciton and −139 meV/% for the B exciton, the latter reaching −180 meV at 1.3% strain. These values are consistent with established biaxial strain coefficients reported for monolayer $WS_2$, where the A exciton typically red-shifts by 130-150 meV/%.[12,44,45] and demonstrate that the accessible strain range in our platform is sufficient to sweep the excitons across the resonance window of the 532 nm excitation.

**Strain-driven suppression of double-resonant 2LA(M) scattering**

The strain-induced redshift of the B exciton and the concurrent modulation of the 2LA(M) Raman mode can be understood within a common physical mechanism. Under biaxial tensile strain, lattice expansion modifies interatomic coupling and the electronic band dispersion, leading to a renormalization of the band structure and a redshift of excitonic transitions[46–48]. In contrast, in our measurements the strain-induced shift of the 2LA(M) feature is much smaller than that of the excitonic transitions. The B exciton shifts by ~138 meV/% strain, whereas the 2LA(M) peak position changes by ~1.34 meV/% (≈1.7 meV over the full strain range), indicating that the resonance modulation is dominated by excitonic detuning rather than by phonon-energy variations. Because the 2LA(M) feature arises from a double-resonant Raman process, its intensity depends on the proximity of the laser excitation to intermediate excitonic transitions.[38,49] While electron–phonon coupling determines the intrinsic probability of phonon-mediated intervalley scattering, the magnitude of the Raman enhancement is governed primarily by the resonance condition[50]. Since excitonic energies shift much more strongly with strain than phonon energies, biaxial strain increases the exciton–laser detuning and progressively suppresses the resonance that amplifies the 2LA(M) process. As a consequence, the Raman response evolves from a resonant regime dominated by intermediate excitonic states toward a non-resonant regime that can be described in terms of virtual intermediate states**.** The smooth and linear evolution of phonon frequencies further indicates that strain-induced changes in electron–phonon coupling are secondary**.** The reduction of the 2LA(M) signal therefore primarily reflects the loss of resonance amplification rather than a direct weakening of the underlying coupling strength.

Because the 532 nm excitation lies within the excitonic resonance window at zero strain, the strain-induced red-shift of the B exciton increases the energy mismatch between the exciton and the laser, leading to a progressive reduction of the double-resonant 2LA(M) band (Figure 5(a,b)). The E and $A_1$ modes remain sharp and retain their characteristic symmetry across the entire strain range, confirming that the decrease of the 2LA(M) intensity originates from resonance detuning rather than structural degradation or increased phonon scattering.

Previous studies did not observe a full detuning-driven crossover because the accessible strain ranges were smaller or strain transfer was insufficient.[13,23] In our platform, efficient mechanical coupling and strain levels exceeding 1% shift the exciton energy across the full resonant window of the laser, allowing the detuning effect to be directly quantified. (Figure 5(b))

It is useful to contrast these biaxial-strain results with the behavior expected under uniaxial deformation. Biaxial tensile strain expands the lattice isotropically and induces a relatively uniform renormalization of the band structure, leading to an efficient redshift of excitonic transitions and therefore to a strong tuning of the exciton–laser detuning that governs the double-resonant Raman process. In contrast, uniaxial strain produces anisotropic band modifications and typically shifts the B exciton less effectively for comparable strain magnitudes, requiring larger strains to reach similar detuning levels.[10] Furthermore, uniaxial strain breaks in-plane symmetry, lifting degeneracies and modifying phonon and valley-related properties, which can introduce additional effects beyond simple resonance detuning. Biaxial strain therefore provides a cleaner and more controlled route to isolate the role of excitonic detuning in the modulation of the 2LA(M) response.

**Exciton-mediated resonance model**

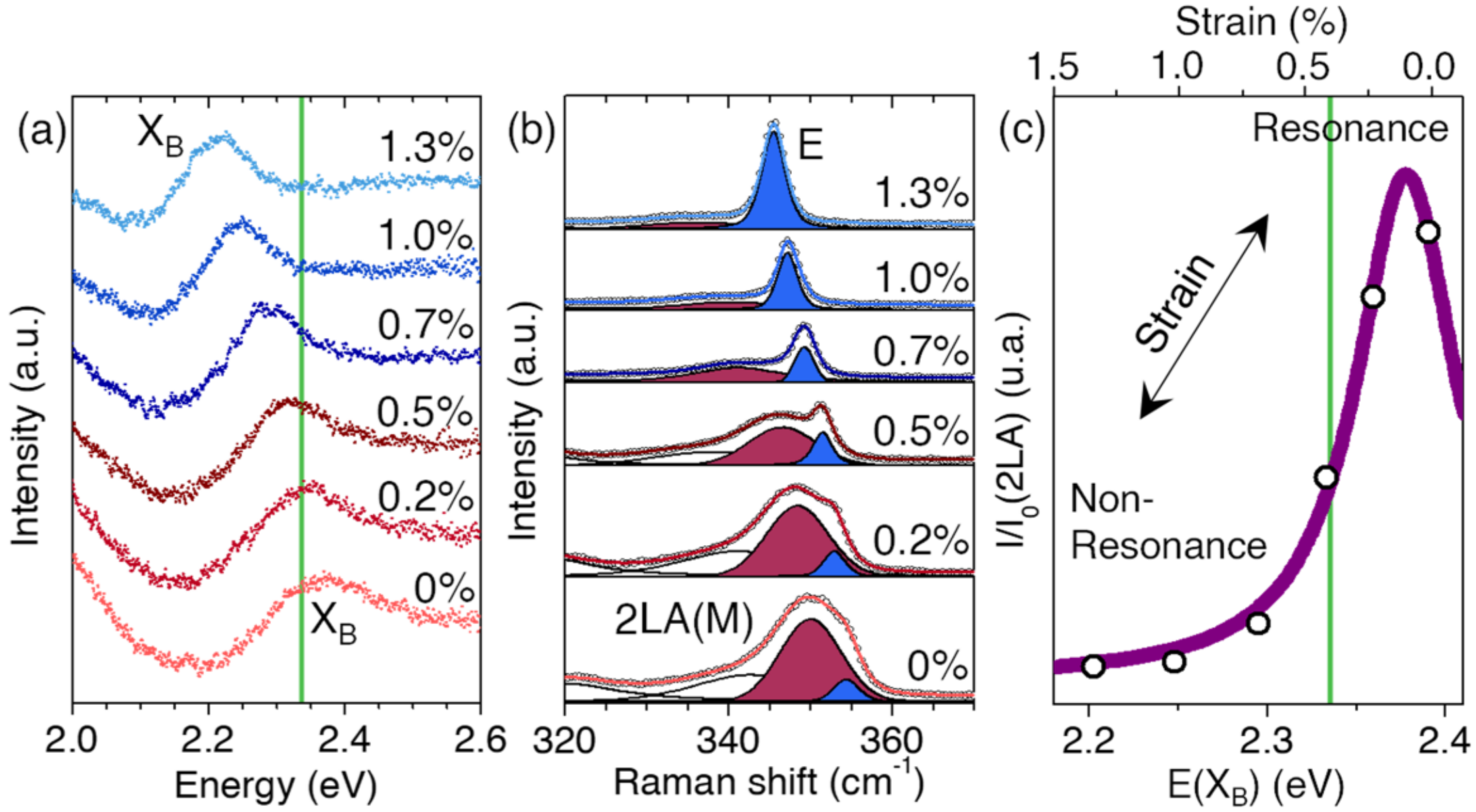

**Figure 5. Strain-controlled transition between resonant and non-resonant 2LA(M) scattering**. (a) B-exciton energy as a function of biaxial strain extracted from differential reflectance. The dashed horizontal line indicates the fixed laser excitation energy (2.33 eV). (b) Strain dependence of the E and 2LA(M) Raman modes positions. (c) Normalized intensity of the 2LA(M) as a function of B-exciton energy, showing a continuous suppression as the exciton is shifted away from resonance with the laser (green line). The upper horizontal axis indicates the corresponding biaxial strain. The purple solid line represents the effective resonance model used to quantify the detuning-dependent decay of the 2LA(M) intensity.

The strain-dependent suppression of the 2LA(M) mode is quantified through its integrated intensity, shown in Figure 5(c), top axis, which directly tracks the progressive reduction of the double-resonant enhancement.

To describe the modulation of the 2LA(M) Raman intensity, we express the resonance directly in terms of the strain-dependent B exciton energy, $E(X_b)$, which mediates the Raman process. In this framework, the relevant quantity is the energy mismatch between the excitation laser energy ($E_L$) and the excitonic intermediate state. Because $E(X_b)$ is experimentally determined at each strain value, the Raman intensity can be written directly as a function of $E(X_b)$ without explicitly invoking strain. To keep the description transparent while capturing the dominant experimental trend, we employ an effective detuning model that describes how the double-resonant enhancement evolves as the excitonic transition is shifted by strain. Within semiclassical exciton-mediated Raman theory, the 2LA(M) intensity is described by[51,52]:

$$I_{2LA(M)} = \frac{A}{(E_L - E(X_b) - \Delta_0)^2 + {\Gamma_{eff}}^2}$$

where A is the exciton-mediated scattering amplitude, $\Gamma_{eff}$ is the effective resonance width incorporating both the intrinsic exciton linewidth and additional broadening introduced by the double-resonant process. Here, $\Delta_0$ is an effective detuning parameter that includes the two-phonon energy contribution. For clarity, we define $\Delta_0 = 2\hbar\omega_{LA} + \delta_0$ where $2\hbar\omega_{LA}$ corresponds to the energy of the two LA phonons involved in the process, and $\delta_0$ accounts for the residual offset between the measured excitonic transition and the effective intermediate state governing the double-resonant pathway. A more general multi-denominator expression derived from double-resonant Raman theory and its reduction to this effective form under exciton-dominated strain tuning are discussed in the Supporting Information.[38,49] This simplified physical picture, where biaxial strain primarily modifies the excitonic detuning while leaving the phonon dispersion only weakly perturbed, allows the resonance modulation of the 2LA(M) intensity to be described directly in terms of the strain-dependent exciton energy.

Fitting the experimental data yields $\Gamma_{eff} \approx 34$ meV and $\Delta_0 \approx -48$ meV. The negative value of $\Delta_0$ indicates that the 2LA(M) intensity is not maximized at exact laser-exciton alignment. Instead, the resonance peaks when the laser lies approximately 50 meV below the B exciton energy (Figure 5(c)), placing the zero-strain excitation on the low-energy shoulder of the resonance. This behaviour is consistent with a phonon-emission-assisted double-resonant mechanism in which the effective intermediate states lie above the real exciton. The extracted $\Gamma_{eff}$ defines the energetic window over which 2LA(M) scattering remains resonantly enhanced. Overall, this formulation demonstrates that the modulation of the 2LA(M) intensity is governed by the absolute exciton energy $E(X_b)$, with strain acting primarily to shift the exciton through the resonance window.

The strain-induced shift of the B exciton and the suppression of the 2LA(M) intensity are reversible. Raman peak positions follow identical loading and unloading trajectories within experimental uncertainty over multiple cycles, with no detectable hysteresis (Supporting Information, Figures S6–S7).

The narrow linewidths of the first-order phonons and the preserved excitonic lineshape confirm that the deformation remains entirely within the elastic regime across the full strain cycle. The ability to reversibly and continuously modulate exciton-phonon interactions demonstrates that biaxial strain acts as a reliable and repeatable control parameter for exciton-mediated optical processes.

Our results show that high biaxial strain can shift the excitonic landscape of trilayer $WS_2$ by ~180 meV, enabling mechanical detuning of resonant Raman scattering without changing the laser energy. This continuous and reversible modulation drives the system into and out of the double-resonant regime and quantitatively reproduces the evolution of the 2LA(M) intensity through an exciton-energy resonance model. The excellent agreement between experiment and theory establishes a direct link between mechanical deformation, excitonic detuning and vibrational scattering. The controlled and quantitative exciton-energy tuning demonstrated here establishes a general strategy for engineering exciton-mediated optical processes using mechanical deformation. By acting as an effective substitute for laser tuning, strain enables deterministic and reversible control of exciton-mediated Raman scattering, opening routes toward mechanically programmable control of resonance-enhanced light–matter interactions in layered semiconductors.

## ASSOCIATED CONTENT

**Supporting Information**.

The Supporting Information includes experimental methods, images of the experimental biaxial setup, AFM images of the samples, spatial uniformity of the applied biaxial strain, additional Raman datasets for monolayer and bilayer $WS_2$, raw data of differential reflectance spectra, experiments demonstrating the reversibility of the applied strain under biaxial strain, and details of the theoretical detuning model. (PDF)

## AUTHOR INFORMATION

Corresponding Author

Álvaro Rodríguez - Instituto de Ciencia de Materiales de Madrid (ICMM)-Consejo Superior de Investigaciones Científicas (CSIC), C. Sor Juana Inés de la Cruz, 3, 28049 Madrid, Spain; https://orcid.org/0000-0003-3703-6712; Email: alvaro.rodriguez@csic.es

**Present Addresses**

† Álvaro Rodríguez- Departamento de Física de la Materia Condensada, Universidad Autonóma de Madrid, Madrid 28049, Spain and Condensed Matter Physics Center (IFIMAC), Universidad Autonóma De Madrid, Madrid 28049, Spain.

**Author Contributions**

A.R. conceived the study, fabricated the samples, carried out the experiments, and performed the data analysis. C.M. and A.C.-G. supervised the project and provided guidance on data interpretation and conceptual development. A.R. wrote the manuscript with input and critical feedback from all co-authors. All authors approved the final version of the manuscript.

**Funding Sources**

A.R acknowledges funding from the European Union under the Marie Skłodowska-Curie Grant Agreement No.101109987. A.C-G. and C.M. acknowledge support from Grants PDC2023-145920-I00 and PID2023-151946OB-I00, funded by MICIU/AEI/10.13039/501100011033 and, respectively, by the European Union NextGenerationEU/PRTR (PDC2023-145920-I00) and by ERDF/EU (PID2023-151946OB-I00). A.C.-G. also acknowledge funding from the European Research Council (ERC) through the ERC-PoC 2024 StEnSo project (grant agreement 101185235) and the ERC-2024 SyG SKIN2DTRONICS project (grant agreement 101167218). ICMM-CSIC authors acknowledge support from the Severo Ochoa Centres of Excellence program through Grant CEX2024-001445-S, funded by MICIU/AEI/10.13039/501100011033.